\documentclass[conference, twocolumn]{IEEEtran}

\usepackage{amsmath,amssymb,graphicx,epsfig,fancyhdr,amsthm,epstopdf}
\usepackage{cite}
\usepackage{amsmath,amssymb,amsfonts,amsthm,bm}
\usepackage{algorithm}%algorithmic   algorithm
\usepackage{algorithmic}
\usepackage{graphicx}
\usepackage{textcomp}
\usepackage{xcolor}
\usepackage{subfigure}
\usepackage{caption, tabularx}

\def\BibTeX{{\rm B\kern-.05em{\sc i\kern-.025em b}\kern-.08em
		T\kern-.1667em\lower.7ex\hbox{E}\kern-.125emX}}
\begin{document}
	
	% paper title
	\title{Deep Learning based Precoding for the \\MIMO Gaussian Wiretap 
	Channel}
	% optimal precoding for MIMO Gaussian Broadcast channels with 
	%integratd Services
	\author{\IEEEauthorblockN{Xinliang Zhang and Mojtaba Vaezi}
%			\IEEEmembership{Senior Member, IEEE}}\\
		\IEEEauthorblockA{Department of Electrical and Computer 
		Engineering, Villanova University}
		%McGill University, Montreal, Quebec H3A 2A7, Canada\\
		Email: \{xzhang4, mvaezi\}@villanova.edu
	}

	% make the title area
	\maketitle
	
	%----------------------------------------------------------------------
	
	\begin{abstract}
		
		A novel precoding method based on supervised deep 
		neural networks  is introduced for the multiple-input 
		multiple-output Gaussian wiretap channel.
		%The scenario that the transmitter delivers confidential messages to the legitimate receiver while keeping eavesdropper perfectly secret from the private messages is considered.
		The proposed deep learning (DL)-based precoding  learns the  input 
		covariance matrix  through offline training over a large set of input 
		channels 
		and their corresponding covariance matrices for   efficient, reliable, and secure 
		transmission of information.    
		%The proposed  DNN network {\color{red}converges quickly} in %regressing the parameters of the beamforming matrix.
		Furthermore, by spending time in  offline training, 
		this method remarkably reduces the computation complexity 
		in real-time applications. 
		%From the results, the activation function and non-linear features can be promising in the DNN-based beamforming method.\\
		Compared to traditional precoding methods, the proposed DL-based 
		precoding  is significantly faster and reaches near-capacity  secrecy 
		rates.  DL-based 
		precoding is also more robust than transitional precoding approaches to  the number of antennas at the  eavesdropper.  This new approach to  precoding is promising in applications  in which   delay and complexity are critical.

	\end{abstract}
	
	%\vspace{-3.5mm}
	%%%%%%%%%%%%%%%%%%%%%%%%%%%%%%%%%%%%%%%%%%%%%%%%%%%%%%%%%%%%%%%%%%%%%%%%%%%%%%%%%%%%%%%%%%%%%%%%%%

	%%%%%%%%%%%%%%%
	% sec III
	%%%%%%%%%%%%%%%
	\section{introduction} \label{sec:intro}
	
	%In this work, we exploit DL not only for a reliable but also for a secure encoding in the context of multiple-input multiple-output (MIMO) Gaussian wiretap channel.
	Wiretap channel \cite{wyner1975wire, csiszar1978broadcast} is a 
	three-node network, consisting of a transmitter, a legitimate receiver, and  an eavesdropper, in which 
	encoding is designed to transmit the legitimate receiver's message  securely and  reliably.  
	This model, which lays the foundation of physical layer security, is then  
	extended to  multi-antenna nodes. The capacity  of multiple-input 
	multiple-output (MIMO) Gaussian wiretap channel  under an average 
	power constraint is established in 
	\cite{khisti2010secure,oggier2011secrecy,liu2009note}. 
	This  capacity expression is abstracted as a non-convex optimization 
	problem over the  covariance matrix of the input signal. This problem is 
	fundamental in the study of physical layer security in the MIMO settings 
	and thus has attracted extensive research in the past decade and has been 
	explored in different ways. Despite this,    the closed-form 
	covariance matrix   is known  only in some special cases 
	\cite{fakoorian2013full, parada2005secrecy,vaezi2017journal}, and optimal 
	signaling to achieve the capacity  is still unknown, in general. 
	
	There are several sub-optimal and iterative solutions for this problem.  \textit{Generalized singular value decomposition} (GSVD)-based precoding,  which splits the transmit channel into several parallel subchannels, provides a closed-form solution \cite{khisti2010secure, fakoorian2012optimal}. 
	This closed-form  solution is, however,  far from capacity in some antenna 
	settings, e.g., when the legitimate receiver has a single antenna \cite{vaezi2017journal}. Alternating optimization and water filling  (AO-WF) algorithm 	\cite{li2013transmit} is another well-known %sub-optimal 
	solution which converts the non-convex problem to a convex problem  and solves 
	it in an iterative manner. However, the complexity  of this method is high 
	and the solution is not stable under certain antenna settings 
	\cite{zhang2019mimome}. {Recently, a new 
	parameterization of the covariance matrix was proposed for two-antenna 
	transmitters  and its optimal closed-form solution was obtained in 
	\cite{vaezi2017journal}. 
	This method is then extended  to arbitrary antennas in
	\cite{zhang2019mimome}. Although the new reformulation of the problem 
	based on the rotation matrices is optimal, the way to find the parameters 
	is iterative and time-consuming, especially for large number of transmit antennas.} Overall, 
	despite extensive research and fundamental importance of this problem, 
	the existing signaling methods, except for closed-form solutions,  suffer 
	either from a high computational complexity or  performance loss.

	Motivate by the above shortcomings and recent successful applications of  
	deep learning (DL)
	in communication over the physical layer \cite{o2017introduction}, in this 
	work, we exploit DL for a 
	secure and reliable signaling design in the MIMO Gaussian wiretap 
	channel. DL is
	a new emerging sub-field of machine learning (ML), and similar to ML,  provides a data-driven approach to  tackle
	traditionally challenging problems \cite{deng2014deep}.
	It holds promise for performance improvements
	in complex  scenarios that are difficult to
	describe with tractable mathematical models or solutions.
	While DL prevails in %certain  fields such as  
	computer vision, speech and audio 
	processing, and natural language processing,
	its introduction to communication systems is relatively new.
	Nonetheless, DL is increasingly being used to solve  communication 
	problems in the physical layer.  Particularly, DL is  being applied to typically 
	hard and intractable problems such as encoding and decoding schemes,  
	beamforming, and power allocation in MIMO, etc. 
	\cite{zappone2019wireless, 
	o2017introduction,zhang2019deep,vaezi2019interplay,besser2019flexible}.

	Recent research efforts have shown that DL is  useful in many  sophisticated communications problems.
	In  \cite{o2017introduction},  a neural network (NN)-based  autoencoder  is 
	proposed for end-to-end reconstruction and communication system 
	design. Although the above work is limited to a differentiable channel,  
	\cite{dorner2017deep} shows that it is possible for autoencoders to work 
	well over the air. Supervised reinforcement learning
	is proposed to characterize communication architecture with a 
	mathematical channel model absence in  \cite{aoudia2018end}. In a more 
	relevant paper to our work, 
	\cite{fritschek2018deep} proposes learning encoding and decoding 
	schemes by NN to realize confidential message transmission over the Gaussian 
	wiretap channel. These successful examples applying DL to 
	communication systems mainly exploit the classification ability of the DL.
	
	In this paper, we develop a DL-based precoding using a \textit{residual 
	network} 
	 \cite{lecun2015deep}  to  reliably and securely transmit 
	information over the MIMO wiretap  channel with near-capacity rates.
	With multiple hidden  and intermediate layers of 
	neurons, the proposed  DL-based precoding  can effectively characterize the precoding 
	matrix. %without knowing the mathematical model of the communication 
	%system. 
	Multiple hidden layers and non-linearity properties of the 
	\textit{deep neural networks} (DNN), enables the proposed DL-based precoding to learn sophisticated mapping 
	 between inputs (channel matrices) and output (covariance matrix). 
	 %which first takes  advantage of the 
	%regression ability of DL.
	Via offline training, the  proposed DL-based precoding learns from a 
	large number of near-optimal 
	covariance matrices and performs well in fitting, regressing and predicting  
	precoding and power allocation matrices.
	Once trained well, the network  achieves a near-capacity secure rate 
	very quickly and with a little memory.  
	Hence, this approach is promising to be applied into Internet of things (IoT), which intrinsically have limited computation abilities and battery life.	
	
	The performance of the proposed precoding, in terms of complexity and 
	achievable rate, is compared with the exiting analytical and 
	numerical solutions, namely, GSVD and  AO-WF. It is shown that similar or better 
	performance can be achieved significantly faster. Moreover, unlike  
	existing solutions, 
	the proposed DL-based precoding is {robust} to the change in 
	the number of antennas at the eavesdropper. This is  meaningful progress 
	towards secure communication in a more practical setting in which the 
	number of antennas at the eavesdropper is unknown.    
	
	The remainder of this paper is organized as follows. In Section~\ref{sec:model}, the system 
	model of the MIMO wiretap channel is described. In 
	Section~\ref{sec:DLnet}, a novel DNN is designed to learn the input 
	covariance matrix.
	The training phase and experimental results are discussed in Section~\ref{sec:trainResult}, and  the  paper is concluded in Section~\ref{sec:conclu}.

	%%%%%%%%%%%%%%%
	% sec II
	%%%%%%%%%%%%%%%

	\begin{figure}[tbp]
		\centering
		\includegraphics[width=0.5\textwidth]{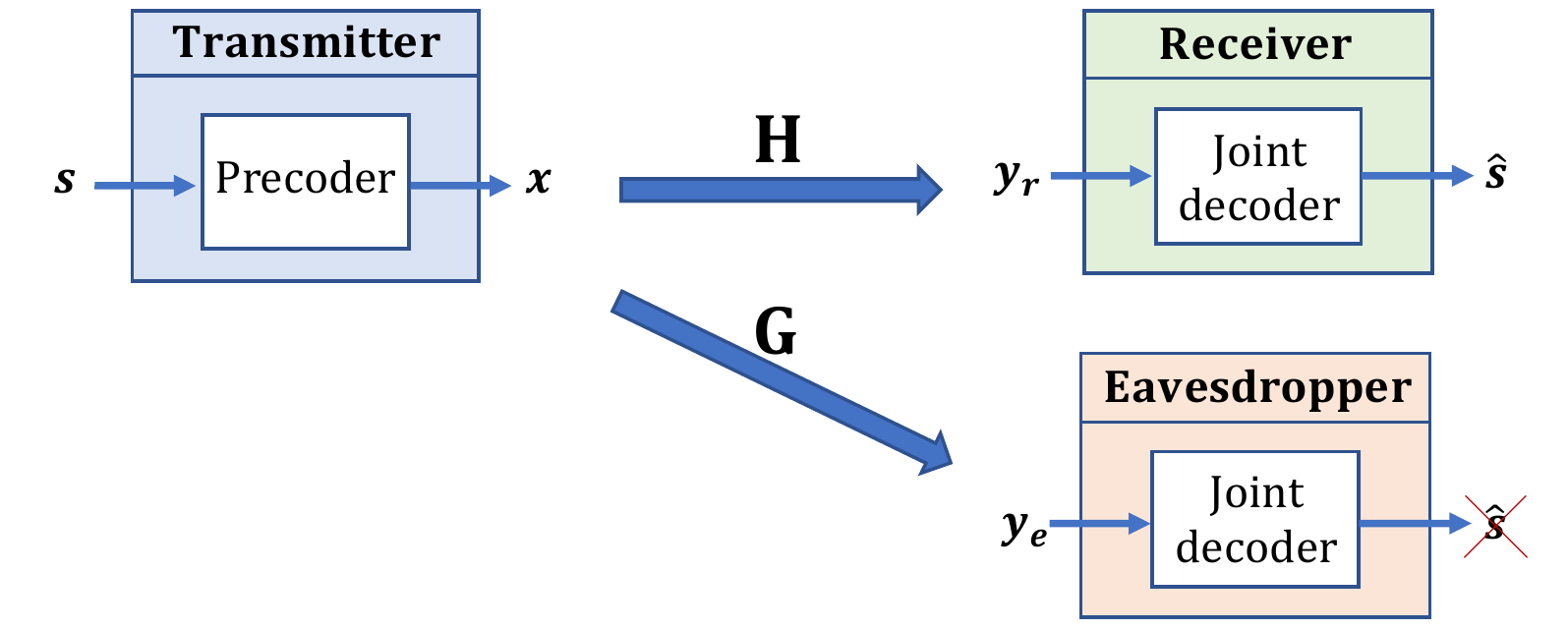}
		% where an .eps filename suffix will be assumed under latex,
		% and a .pdf suffix will be assumed for pdflatex
		\caption{The MIMO wiretap  channel.}
		\label{fig:WiretapModel}
	\end{figure}
	
	\section{System Model} \label{sec:model}
	The MIMO wiretap  channel is  a model for \textit{reliable} and  \textit{secure} communication over the air 	which includes a transmitter equipped with $n_t$ antennas which sends 
	a   message to a  legitimate receiver  with $n_r$ 
	antennas while keeping it confidential  from an   eavesdropper 
	equipped with $n_e$ antennas. 
	The system model is shown in Fig.~\ref{fig:WiretapModel} in which $\mathbf{s} \in 
	\mathbb{R}^{n_t}$ is the information vector, $\mathbf{x} \in 
	\mathbb{R}^{n_t}$ is the transmitted signal, $\mathbf{y}_r \in 
	\mathbb{R}^{n_r}$ and $\mathbf{y}_e \in 
	\mathbb{R}^{n_e}$ are the received signal at receiver and eavesdropper 
	sides. The received signals at the legitimate receiver and the eavesdropper 
	sides at time 
	$m$
	can be, respectively, expressed as
	%\begin{align}\label{eq:recSig_h}
	%	\mathbf{y}_r = \mathbf{Hx} + \mathbf{w}_r,\quad\mathbf{y}_e = 
	%	\mathbf{Gx} + 
	%	\mathbf{w}_e,
	%\end{align}
	\begin{subequations} \label{eq:recSig_h}
		\begin{align} 
		%\mathbf{Y}_r[m] = \mathbf{H}\mathbf{X}[m] +\mathbf{W}_r[m],\\
		%\mathbf{Y}_e[m] = \mathbf{G}\mathbf{X}[m] +\mathbf{W}_e[m],
		\mathbf{y}_r[m]   = \mathbf{H}\mathbf{x}[m] +\mathbf{w}_r[m] ,\\
		\mathbf{y}_e[m] = \mathbf{G}\mathbf{x}[m]   +\mathbf{w}_e[m] ,
		\end{align}
	\end{subequations}
	in which $\mathbf{H} \in \mathbb{R}^{n_r \times n_t}$  and $\mathbf{G} 
	\in \mathbb{R}^{n_e \times n_t}$ are the  channel matrices 
	corresponding to the 
	receiver and eavesdropper,  $\mathbf{w_r} \in \mathbb{R}^{n_r}$ 
	and $\mathbf{w_e}\in \mathbb{R}^{n_e}$ are Gaussian white noises with zero 
	means and identity covariance matrices. 
	The channel input is subject to an average total power constraint 
	\cite{liu2009note}, i.e.,
	\begin{align} \label{eq:power cons}
	%\frac{1}{n}\sum_{m=1}^{n}||\mathbf{X}||^2 = 
	%\frac{1}{n}\sum_{m=1}^{n}(\mathbf{X}[m]^{T}\mathbf{X}[m]) \leq P
	%\frac{1}{M}\sum_{m=0}^{M-1}(||\mathbf{x}[m]||^2) = 
	||\mathbf{x}||^2 = 
	\frac{1}{M}\sum_{m=0}^{M-1}(\mathbf{x}[m]^{T}\mathbf{x}[m]) \leq P,
	\end{align}
	where $M$ is the  length of $\mathbf{x}$. 
%	Csisz\'ar and K\"orner 
%	\cite{csiszar1978broadcast} provided a 
%	information-theoretic expression for secrecy capacity 
%	region of the discrete memoryless wiretap channel which is given by
%	\begin{align} \label{eq:inforregion}
%	\mathcal{C}_s & = \max \limits_{p(x)} [I(U;Y_r) -  I(U;Y_e)],
%	\end{align}
%	in which $U$ is an auxiliary random variable, $p(x)$ is the probability 
%mass 
%	function of random variable $X$, % and the bold script $\mathbf{x}$ is 
%%the realizations.
%	and  $I(A;B)$ is the mutual 
%	information between random variables $A$ and $B$.
	The capacity expression 
	of the MIMO wiretap  channel \eqref{eq:recSig_h} under the average 
	power 
	\eqref{eq:power cons} is expressed as  \cite{liu2009note}
	\begin{align}\label{eq:Rate}
	\mathcal{C}_s = &\max \limits_{\mathbf{Q}} 
	\frac{1}{2}\log|\mathbf{I}_{n_r}+\mathbf{HQH}^T|-
	\frac{1}{2}\log|\mathbf{I}_{n_e}+\mathbf{GQG}^T|, \nonumber \\
	&\;\;\mathrm{s.t.} \quad\mathbf{Q}\succeq 0, 
	\textmd{tr}(\mathbf{Q}) \leq P,
	\end{align}
	in which the  covariance matrix $\mathbf{Q}=E\{\mathbf{xx}^T\} \in 
	\mathbb{R}^{n_t 
		\times n_t}$ is symmetric and positive semi-definite, and  
		$\mathbf{A}^T$, 
	$\textmd{tr}(\mathbf{A})$, 
	and $|\mathbf{A}|$ represent transpose, trace, and  determinant of matrix $\mathbf{A}$, respectively.
	
	Optimal closed-form $\mathbf{Q}$ is known only for special numbers of antenna 
	settings 
	\cite{zhang2019mimome}.  There, however, are a few well-known sub-optimal analytical 
	and numerical  solutions for arbitrary numbers of antennas. 
	Among them are GSVD, AO-WF, and rotation-based precoding, as 
	discussed earlier. We note that since eigenvalue decomposition 
	of $\mathbf{Q}$ results in $	\mathbf {Q} =\mathbf{V} \mathbf {\Lambda } 
	\mathbf{V}^t$, we can design the transmitted signal vector as
	\begin{align}\label{eq:lp}
	\mathbf{x}=\mathbf{V} \mathbf{\Lambda}^{\frac{1}{2}}   \mathbf{s},
	\end{align}
	in which 
	\begin{itemize}
		\item $\mathbf{V}$   is the \textit{precoding matrix},	
		\item $\mathbf{\Lambda}^{\frac{1}{2}}$  is the  \textit{power allocation matrix}, and	
		\item $\mathbf{s} $ is the information signal vector with covariance  $\mathbf{I}$.
	\end{itemize} 
	Thus, knowing the covariance matrix is equivalent to knowing the 
	corresponding precoding and power allocation matrices. Hence, these two are
	used equivalently in this paper.
%	{\color{red}As a basic information-theoretic model, maximizing the 
%	capacity based on 
%	mutual information from the input samples is introduced in 
%	\cite{barber2003algorithm}, and recently is applied to generative 
%	adversarial 	networks (GANs) \cite{chen2016infogan}. The optimization 
%	fashion based on 	information entropy can also be realized in 
%	\cite{fritschek2018deep} through 	NN. From a signal processing point of 
%	view, we design a NN as a precoding 	scheme to reach the capacity.}

	%%%%%%%%%%%%%%%
	% sec III
	%%%%%%%%%%%%%%%
	\section{Deep Learning Architecture for Precoding}\label{sec:DLnet}
	This paper designs a precoder based on a DNN for 
	the MIMO Gaussian wiretap 
	channel. The inputs of the network are  
	channel matrices 
	 $ (\mathbf{H} $ and $ \mathbf{G})$ and their non-linear combinations. 
	The output 
	is the upper triangular part of the optimal covariance matrix 
	$\mathbf{Q}$.  After the training process, the network learns the features of 
	optimal 
	covariance 
	matrix $\mathbf{Q}$ over different channels. In this paper, $ \mathbf{Q} $ used for training is
	obtained from AO-WF method and is called optimal $ \mathbf{Q}$.  In fact, the 
	network tries to learn how to get a $ \mathbf{Q} $ similar to that of AO-WF.
Alternatively,  rotation-based method \cite{zhang2019mimome} can be used.

	\subsection{Input Features}	\label{sec:input}
\begin{figure}[!t]
		\centering
		\includegraphics[width=0.46\textwidth]{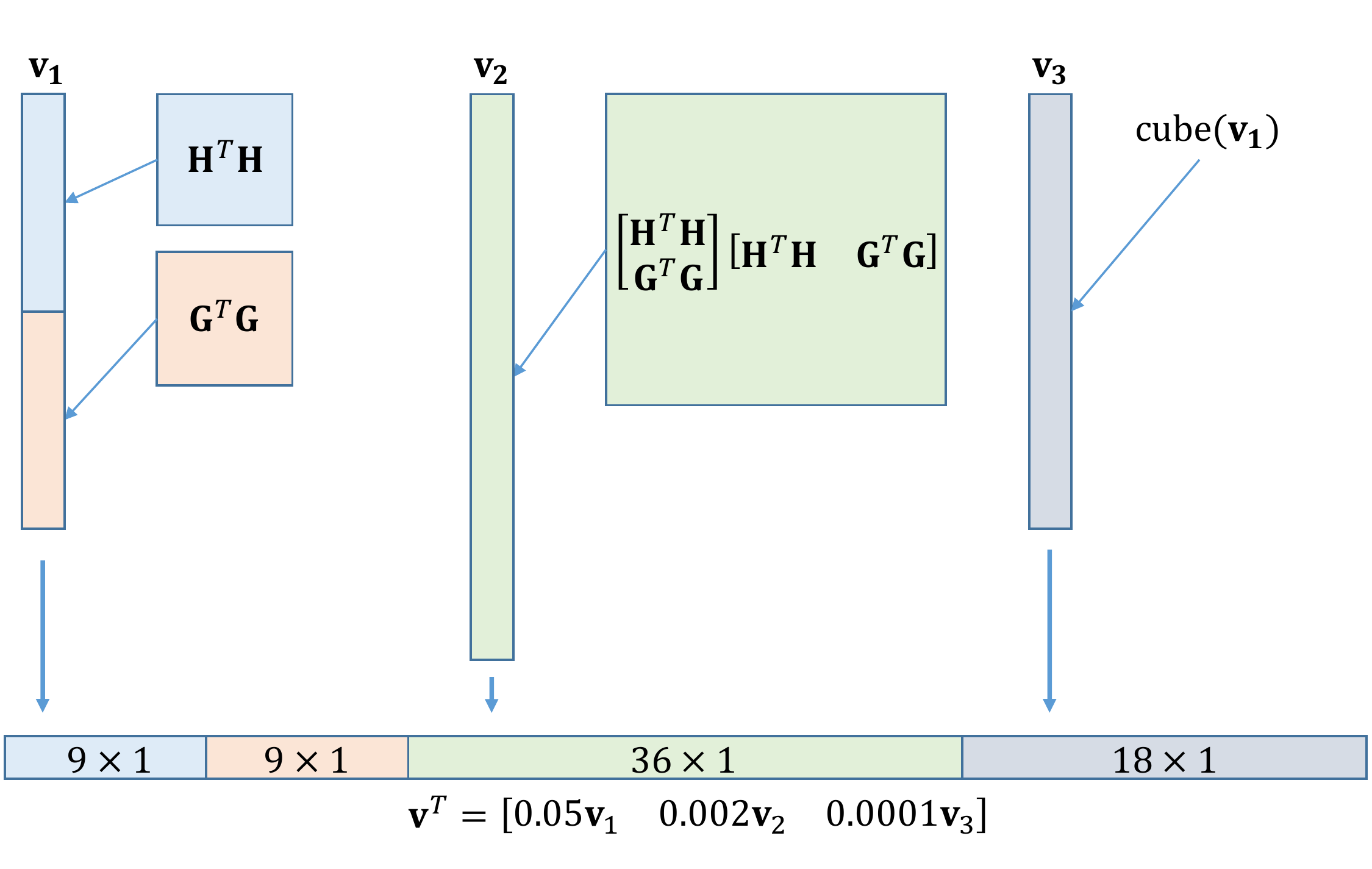}
		\caption{The input design.}
		\label{fig:inputs}
	\end{figure}

	For $n_t=2$, optimal $ \mathbf{Q} $ is known analytically from 
	\cite{vaezi2017journal}. Here we consider $n_t=3$ in this paper\footnote{Without loss 
	of generality, the network with arbitrary $n_t$ can be realized by changing 
	the size of inputs.}. The network  input vector $\mathbf{v}$ contains 72 
	features as shown in Fig.~\ref{fig:inputs}.  
	These features include the elements of channel matrices, i.e.,  $\mathbf{H}$ 
	and 
	$\mathbf{G}$, and their non-linear combinations as shown in 
	Fig.~\ref{fig:inputs}. Note that sing Sylvester's 
	determinant identity  the arguments of the logarithms in \eqref{eq:Rate}  can be written as 
	\begin{subequations}
		\begin{align}
		\left|
		\mathbf{I}_{n_r}+\mathbf{H}\mathbf{Q}\mathbf{H}^T
		\right|=\left|
		\mathbf{I}_{n_t}+\mathbf{H}^T\mathbf{H}\mathbf{Q}
		\right|,   \label{eq:detAround_a}\\
		\left|
		\mathbf{I}_{n_e}+\mathbf{G}\mathbf{Q}\mathbf{G}^T
		\right|=\left|
		\mathbf{I}_{n_t}+\mathbf{G}^T\mathbf{G}\mathbf{Q}
		\right|.   \label{eq:detAround_b}
		\end{align}
	\end{subequations}
Hence, 	$\mathbf{H}^T\mathbf{H}$ or $\mathbf{G}^T\mathbf{G}$ can be 
	considered as a whole which are both $n_t \times n_t$ matrices. In this paper, the input vector $\mathbf{v}$ is 
	designed as
	\begin{align}\label{eq:inputVec}
	\mathbf{v}\triangleq[0.05\mathbf{v}_1, 
	0.002\mathbf{v}_2,0.0001\mathbf{v}_3]^T,
	\end{align}
	in which $\mathbf{v}_1$,  $\mathbf{v}_2$, and  $\mathbf{v}_3$ are 
	defined as 
	\begin{subequations}
		\begin{align}
		&\mathbf{v}_1\triangleq        {\rm vec}([\mathbf{H}^T\mathbf{H} \;\;
		\mathbf{G}^T\mathbf{G}]), \label{eq:inputV1}\\
		&\mathbf{v}_2\triangleq        {\rm vec}([\mathbf{H}^T\mathbf{H} \;\;
		\mathbf{G}^T\mathbf{G}]^T[\mathbf{H}^T\mathbf{H} \;\;
		\mathbf{G}^T\mathbf{G}]), \label{eq:inputV2}\\
		&\mathbf{v}_3\triangleq {\rm cube}( \mathbf{v}_1), \label{eq:inputV3}
		\end{align}
	\end{subequations}
	where ${\rm vec}(\cdot)$ is the vectorization of a matrix and ${\rm 
		cube}(\cdot)$ is the element-wise cube operation. The coefficients of these 
	vectors are used for normalization and 
	weighting.  The sketch of the  input 
	vector is shown in 	Fig.~\ref{fig:inputs}. Here, {$ \mathbf{v}_1 $ 
		can be 
		seen as an original feature  whereas   $ \mathbf{v}_2 $ and $ \mathbf{v}_3 
		$ 
		provide additional nonlinear combination of the original features which 
		increases the probability that DNN can learn the mapping from input to 
		desired 
		output.}

	\subsection{Network Design}\label{sec:network}
\begin{figure}[t]
		\centering
		\includegraphics[width=0.43\textwidth]{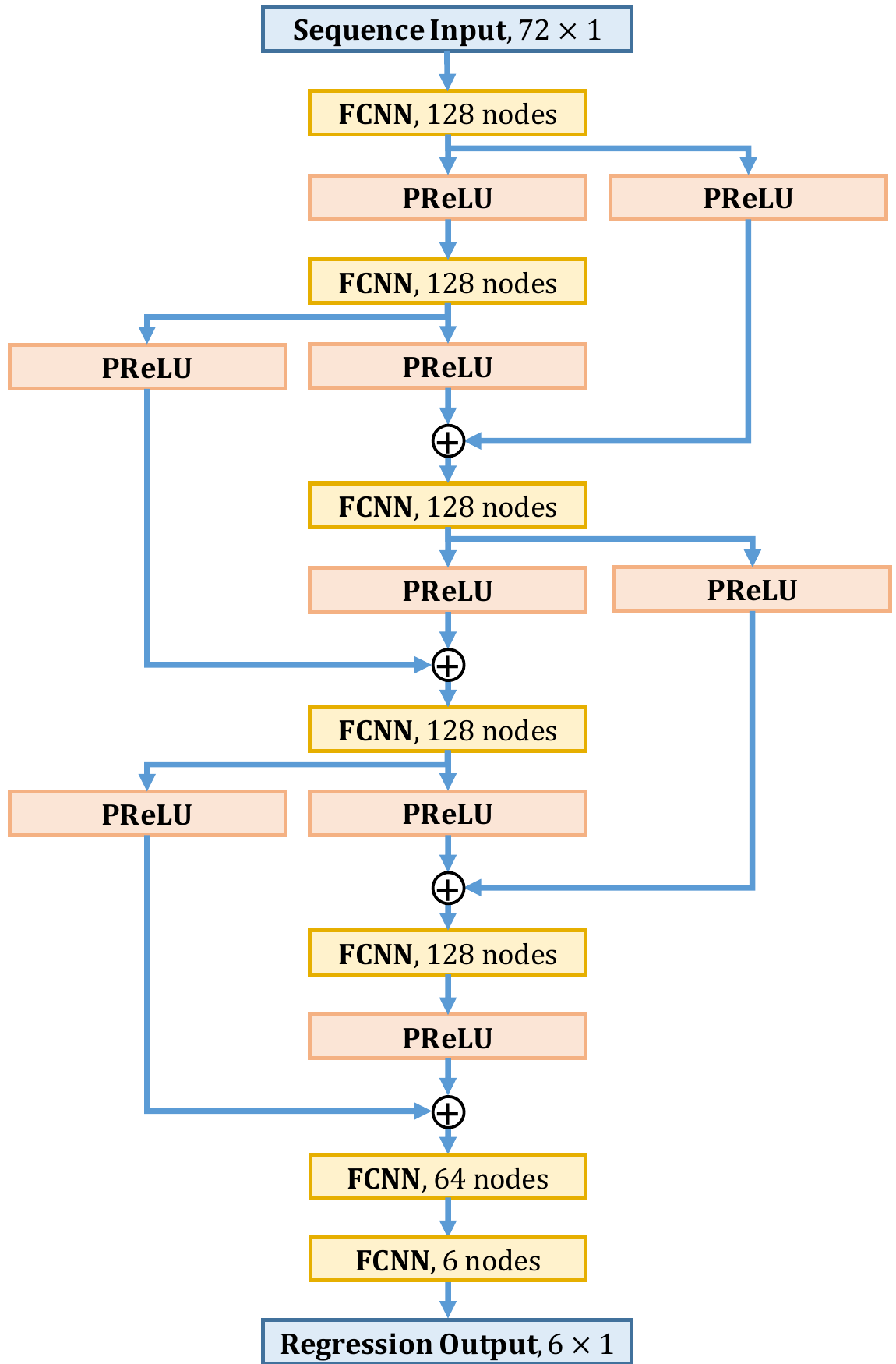}
		\caption{The proposed DL architecture.}
		\label{fig:network}
	\end{figure}

	The network architecture for $n_t=3$  is shown in Fig.~\ref{fig:network}. This is a  
	\textit{fully-connected  
	 neural network} (FCNN) with \textit{parametric 
	 rectified linear units} (PReLUs) 
	\cite{he2015delving} as activation functions.  FCNN can be seen as a special  convolutional neural network with filter size $1\times1$ \cite{long2015fully}. PReLU can provide more 
	trainable 	parameters and prevent over-fitting\cite{he2015delving}. 
	Besides, we introduce a few 
	\textit{shortcut 
		connections} proposed in the \textit{residual network} 
		\cite{he2016deep}. 
	The shortcut connections are able to reduce the difficulty of training 
	process and make the network converge better. We 
	add a PReLU layer with 
	unique trainable parameters to each  shortcut connection.
	%  and the outputs of the 
%	network with two 	fully-convolutional layers gradually decrease the number of hidden 
%	nodes to 	$6$. 

	\subsection{Expected Output}\label{sec:output}
	The covariance matrix for $n_t=3$ is given by
		\begin{align}\label{eq:Q}
	\mathbf{Q}=\left[
	\begin{array}{ccc}
	q_{11} & q_{12} & q_{13}\\
	q_{12} & q_{22} & q_{23}\\
	q_{13} & q_{23} & q_{33}
	\end{array}
	\right].
	\end{align}
	The output vector $\mathbf{q}$ contains the upper triangular part of the 
	covariance matrix $\mathbf{Q}$ since it is symmetry. More specifically, 
	\begin{align}\label{eq:onputVec}
	\mathbf{q}\triangleq[q_{11},q_{22},q_{33},q_{12},q_{23},q_{13}]^T.
	\end{align}
Each  $\mathbf{Q}$ is given by AO-WF \cite{li2013transmit}. The network is 
required to learn the relation between the channels and  $\mathbf{Q}$.

	%%%%%%%%%%%%%%%
	% sec IV
	%%%%%%%%%%%%%%%
	\section{Training Procedure and Simulation Results}\label{sec:trainResult}
	The training procedure and regression results are demonstrated in this 
	section. We also examine the performance of DL-based precoding in this 
	section.
	
	\subsection{Data Set Generation}\label{sec:dataGen}
	The experiments are associated with three training sets, i.e.,  
	\textit{TrainingSet-I}, \textit{TrainingSet-II} and \textit{TrainingSet-III},  as 
	shown in Table~\ref{tab:trainingSet}. 
	\textit{TrainingSet-I} and \textit{TrainingSet-II} contain $2,000,000$ 
	samples. Each sample contains $72$  input features contributed by the  
	channel matrices. The channels are generated randomly following the 
	standard Gaussian distribution. \textit{TrainingSet-I} is for  $n_t=3$, 
	$n_r=4$, and $n_e=3$ whereas in  \textit{TrainingSet-II} the number of 
	antennas are $n_t=3$, $n_r=2$, and $n_e=1$.

	\begin{table}[tbp]
		\caption{Details of the Data Sets.}
		\label{tab:trainingSet}
		\centering
		\begin{tabular}{c|cccc}
			\hline
			& $n_t$      & $n_r$      & $n_e$      & number of samples     \\ 
			\hline
			\textit{TrainingSet-I} & 3          & 2          & 1          & 
			2,000,000             \\
			\textit{TestSet-I}     & 3          & 2          & 1          & 1000                 \\
			\textit{TrainingSet-II} & 3          & 4          & 3          & 
			2,000,000             \\
			\textit{TestSet-II}     & 3          & 4          & 3          & 1000                 
			\\ 
			\hline
			\textit{TrainingSet-III} & \multicolumn{4}{l}{Cascade of 
				\textit{TrainingSet-I} and \textit{TrainingSet-II}} \\
			\textit{TestSet-III}     & \multicolumn{4}{l}{Cascade of \textit{TestSet-I} 
				and 
				\textit{TestSet-II}}         \\ \hline
		\end{tabular}
	\end{table}

	Then, AO-WF \cite{li2013transmit} is used to
	generate optimal $\mathbf{Q}$ for each set of channels and the total 
	average transmit power 
	constraint is $P\leq20W$ for all cases.  The upper triangular part of $\mathbf{Q}$ is  the output used for 
	supervised 
	learning. \textit{TrainingSet-III} is the cascade of \textit{TrainingSet-I} and 
	\textit{TrainingSet-II} with random order of samples. We also generate 
	\textit{TestSet-I} and \textit{TestSet-II} as test data sets, each of which 
	having $1000$ samples with 
	antenna setting corresponding to \textit{TrainingSet-I} and 
	\textit{TrainingSet-II}.

	\subsection{Training Process}\label{sec:training}
	In the training process, the proposed DL-based precoding is trained three times  using 
	\textit{TrainingSet-I}, \textit{TrainingSet-II} and 
	\textit{TrainingSet-III}. The training process is executed on a single graph 
	card (NVIDA GeForce GTX 
	1080) and Adam\cite{kingma2014adam} is used as the optimization method. 
	Except for the 
	batch 
	size, all training process has the same hyper-parameters.  The total epochs are 
	$4,000,000$. Learning rate is initially set $0.001$ and then is 
	decreased $20\%$ after every $80$ epochs. The  batch size  for 
	\textit{TrainingSet-I} and 
	\textit{TrainingSet-II} is $2000$ whereas for \textit{TrainingSet-III} it is 
	$4000$. 
	Considering the number of samples in \textit{TrainingSet-III} is twice 
	as many as that in \textit{TrainingSet-I} and \textit{TrainingSet-II}, the 
	doubled 
	batch size will ensure the training time consumption for all training 
	process was 
	approximately the same; it was about $20$ hours in our experiments.

\begin{figure}[t]
	\centering
	\includegraphics[width=0.45\textwidth]{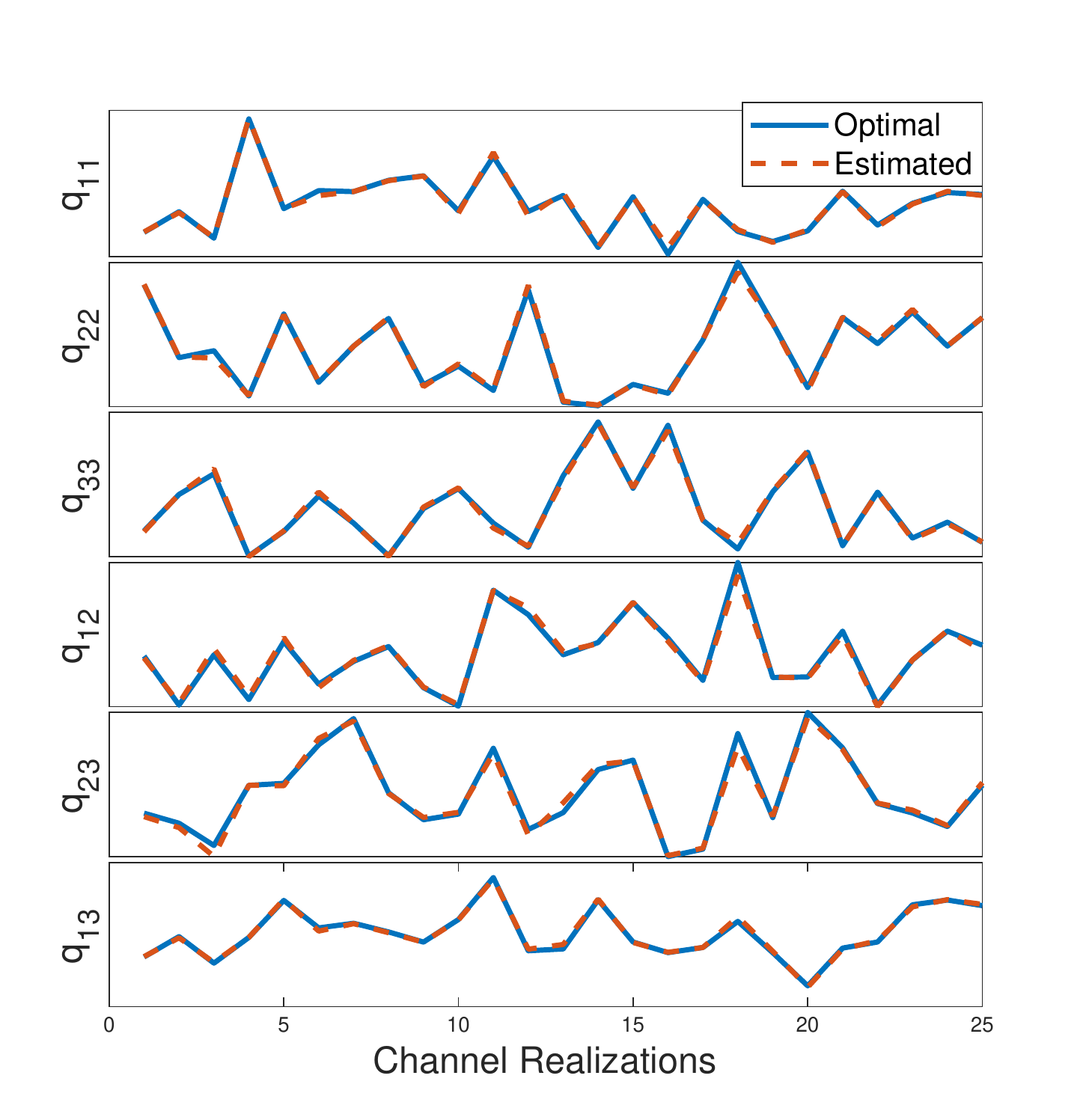}
	\caption{Comparison between optimal and estimated elements in 
		$\mathbf{Q}$ 
		when  
		$n_t=3$, $n_r=2$, and 
		$n_e=1$.}
	\label{fig:QMSE_321}
\end{figure}
	
		\begin{figure*}[t]
		\centering
		\subfigure[\textit{TrainingSet-I} with \textit{TestSet-I}]{
			\includegraphics[width=0.47\textwidth]{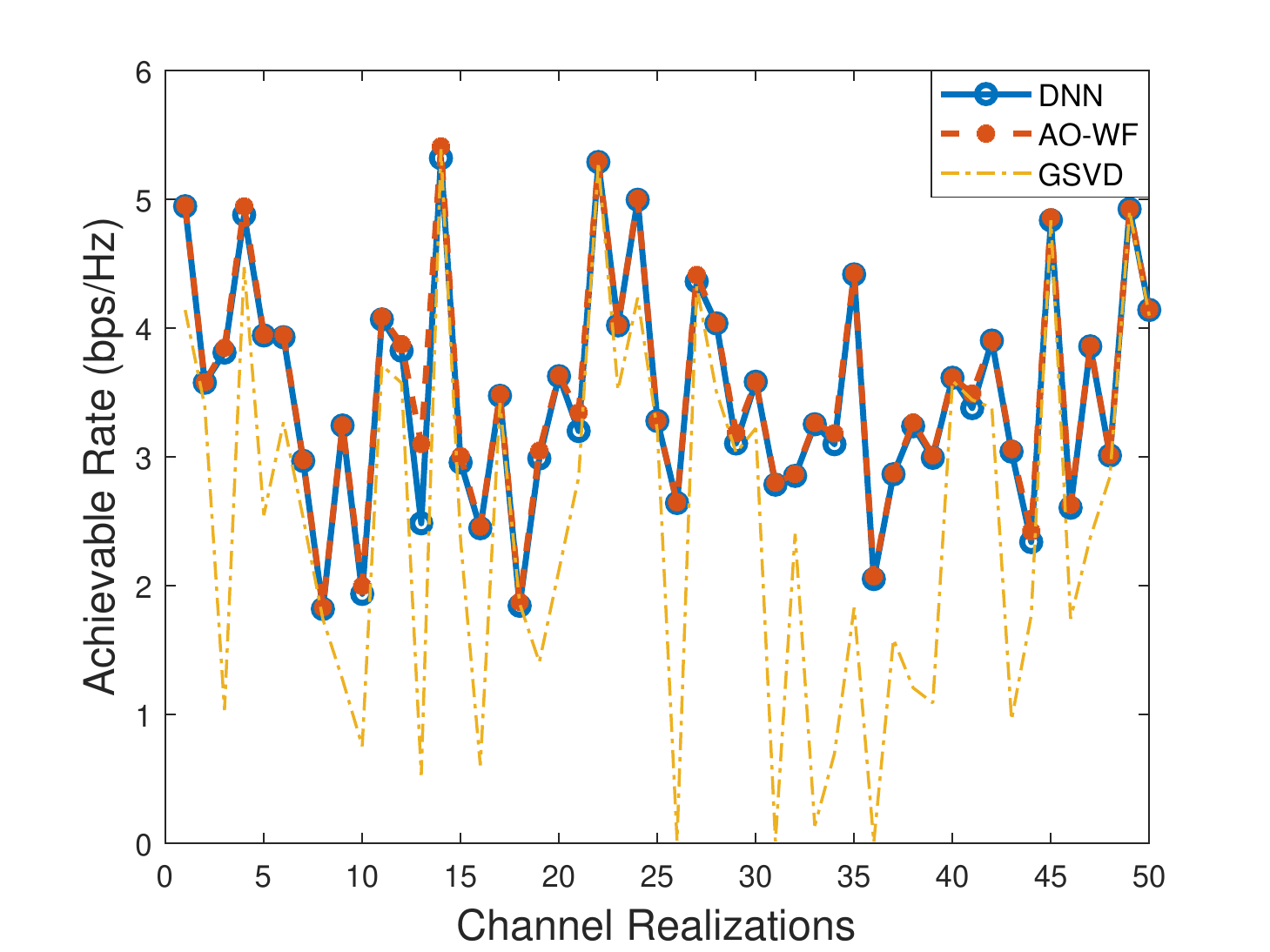}
			\label{fig:AchivaRate_comp_321}}
		\subfigure[\textit{TrainingSet-II} with \textit{TestSet-II}]{
			\includegraphics[width=0.47\textwidth]{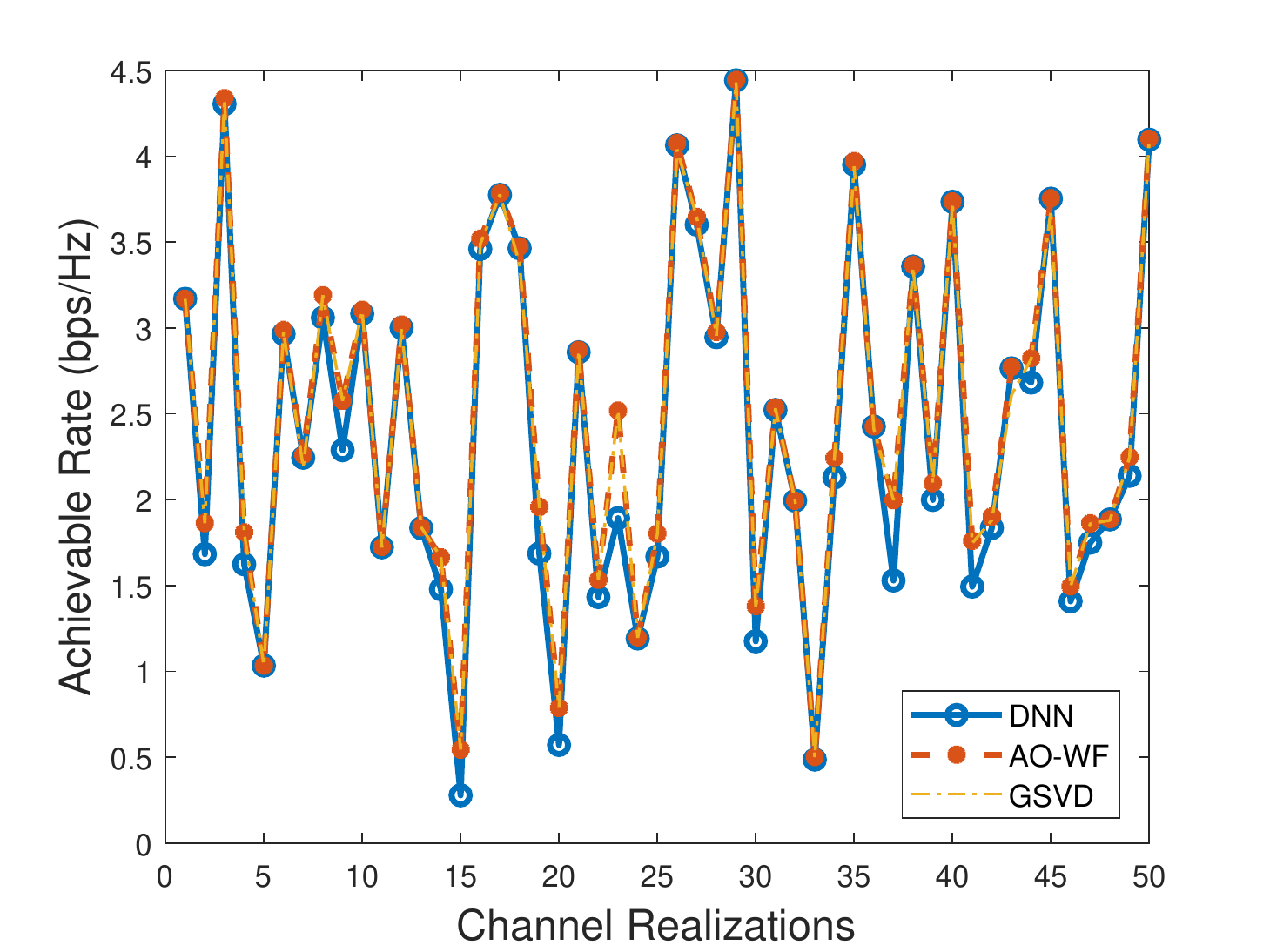}
			\label{fig:AchivaRate_comp_343}}
		\caption{Comparison of achievable secrecy rate using corresponding training and test sets.}
	\end{figure*}

	\subsection{Performance of the DL-based Precoding} 
	\label{sec:result}
	The performance of the proposed DL-based precoding can be evaluated by 
	corresponding training and test data 
	sets, i.e., \textit{TrainingSet-I} with \textit{TestSet-I} and 
	\textit{TrainingSet-II} 
	with \textit{TestSet-II}.  The mean squared error (MSE)  for evaluating
	$\mathbf{Q}$ is shown in Table~\ref{tab:testMSE}. Besides, 
	Fig.~\ref{fig:QMSE_321}
	  illustrates the estimation results and 
	their expected values for \textit{TestSet-I}. It is seen that the elements in 
	$\mathbf{Q}$ are  estimated with fairly good  MSEs.

	Once  the network ``learns'' to estimate the optimal $\mathbf{Q}$, it is 
	ready to be used for precoding and power allocation based on 
	\eqref{eq:lp}.
	The achievable rate versus channel realizations is shown in 
	Figs.~\ref{fig:AchivaRate_comp_321} and \ref{fig:AchivaRate_comp_343}, 
	and is compared with those of AO-WF and GSVD. {Further, the average   
		secrecy rate 
		of each test process  is provided 
		in  Table~\ref{tab:Rcomp}.}
	\begin{table}[htb]
		\caption{MSE of DL-based Precoding with Corresponding Training 
			and Test Data Sets.}
		\label{tab:testMSE}
		\centering
		\begin{tabular}{c|cc}
			\hline
			Training Data Set & 		\textit{TrainingSet-I} & 	
			\textit{TrainingSet-II} \\
			Test Data Set  & \textit{TestSet-I} & 	\textit{TestSet-II} 
			\\ \hline
			MSE of 	$\hat{q}_{11}$ & 0.1402                  & 0.0779                  \\
			MSE of	$\hat{q}_{22}$ & 0.1338                  & 0.0740                  \\
			MSE of	$\hat{q}_{33}$ & 0.1479                  & 0.0633                  \\
			MSE of	$\hat{q}_{12}$ & 0.1167                  & 0.0770                  \\
			MSE of	$\hat{q}_{23}$ & 0.1425                  & 0.0741                  \\
			MSE of	$\hat{q}_{13}$ & 0.1220                  & 0.0711                  \\ 
			\hline
		\end{tabular}
	\end{table}

\begin{table}[htbp]
	\caption{Average  Achievable Secrecy Rate.}
	\label{tab:Rcomp}
	\centering
	\begin{tabular}{cc|ccc}
		\hline
		Training Set & Test Set& DL-based  & AO-WF & GSVD \\ \hline
		\textit{TrainingSet-I} &\textit{TestSet-I} & 3.3980 & 3.4741 & 2.5197 \\
		\textit{TrainingSet-II} &\textit{TestSet-II} & 2.3381 & 2.4827 & 2.4639\\
		\hline
		\textit{TrainingSet-III} &\textit{TestSet-I} &3.3947 & 3.4741 & 2.5197 \\
		\textit{TrainingSet-III} &\textit{TestSet-II} &2.3267 & 2.4827 & 
		2.4639\\ \hline
	\end{tabular}
\end{table}

	As can be seen from the figures, the proposed DL-based precoding is able to reach a  
	secrecy rate comparable to that of  
	AO-WF. Besides, the proposed DL-based precoding performs better than GSVD in the case 
	$n_t=3$, $n_r=2$, and $n_e=1$. 
	More importantly,  as illustrated in 
	Table~\ref{tab:timeCost},  the 
	proposed 
	DL-based precoding is much more 
	efficient than the 
	traditional methods. Although  DL's training time is long, 
	the training process  is usually realized offline.
	Accordingly, a well-trained precoding is a promising tool 
	in the Gaussian MIMO wiretap problem especially for IoT devices with 
	limited energy and computing power.

\begin{table}[htbp]
	\caption{Average Time  for One Realization.}
	\label{tab:timeCost}
	\centering
	\begin{tabular}{c|ccc}
		\hline 
		& DL-based & AO-WF & GSVD\\
		\hline
		Time Cost (ms) & 0.0255 & 243 & 0.513\\ \hline
	\end{tabular}
\end{table}

	\subsection{Cascading Training Sets } \label{sec:cascad}
\begin{figure*}[t]
	\centering
	\subfigure[\textit{TrainingSet-III} with \textit{TestSet-I}]{
		\includegraphics[width=0.47\textwidth]{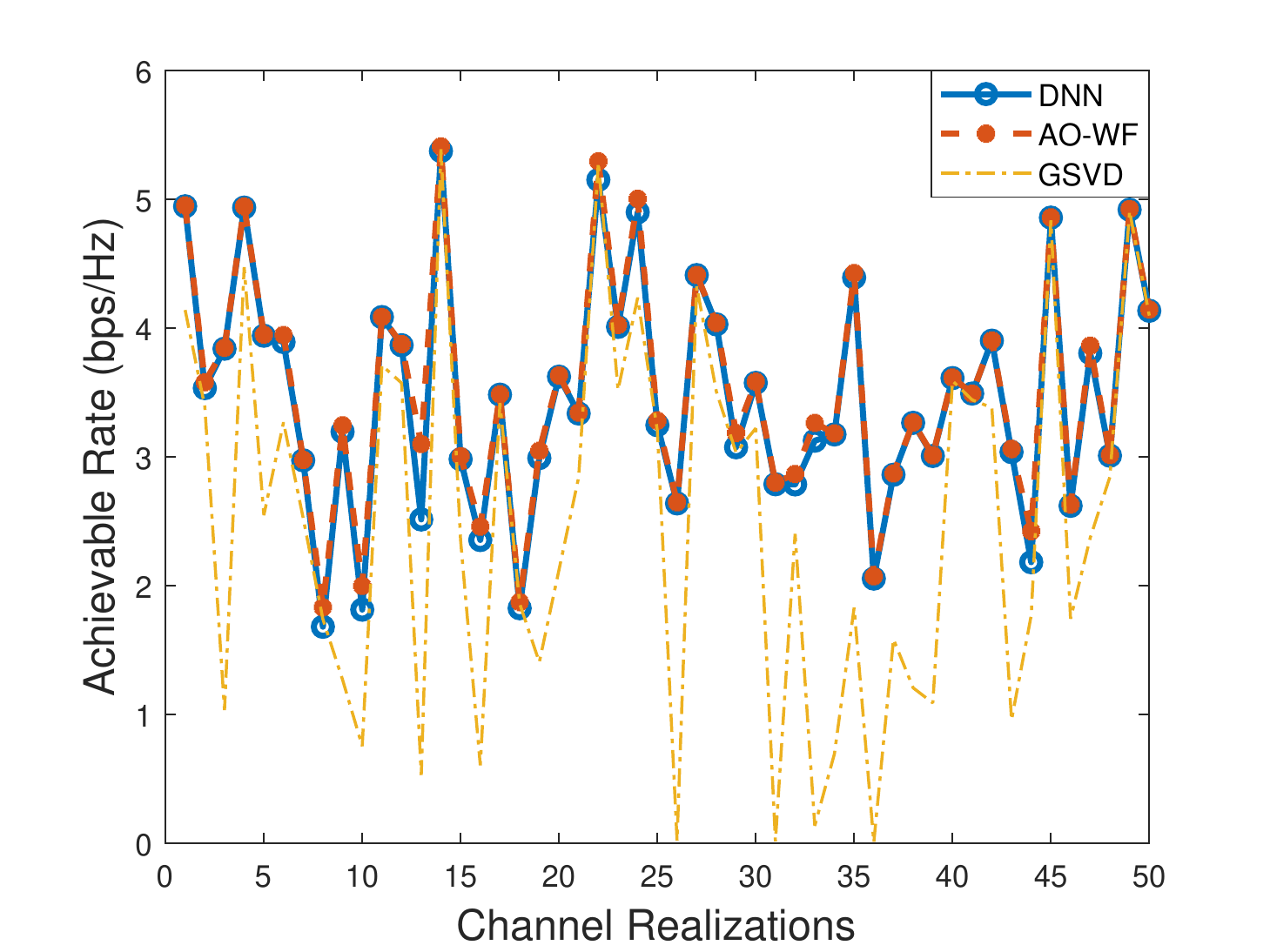}
		\label{fig:AchivaRate_jointly_321}}
	\subfigure[\textit{TrainingSet-III} with \textit{TestSet-II}]{
		\includegraphics[width=0.47\textwidth]{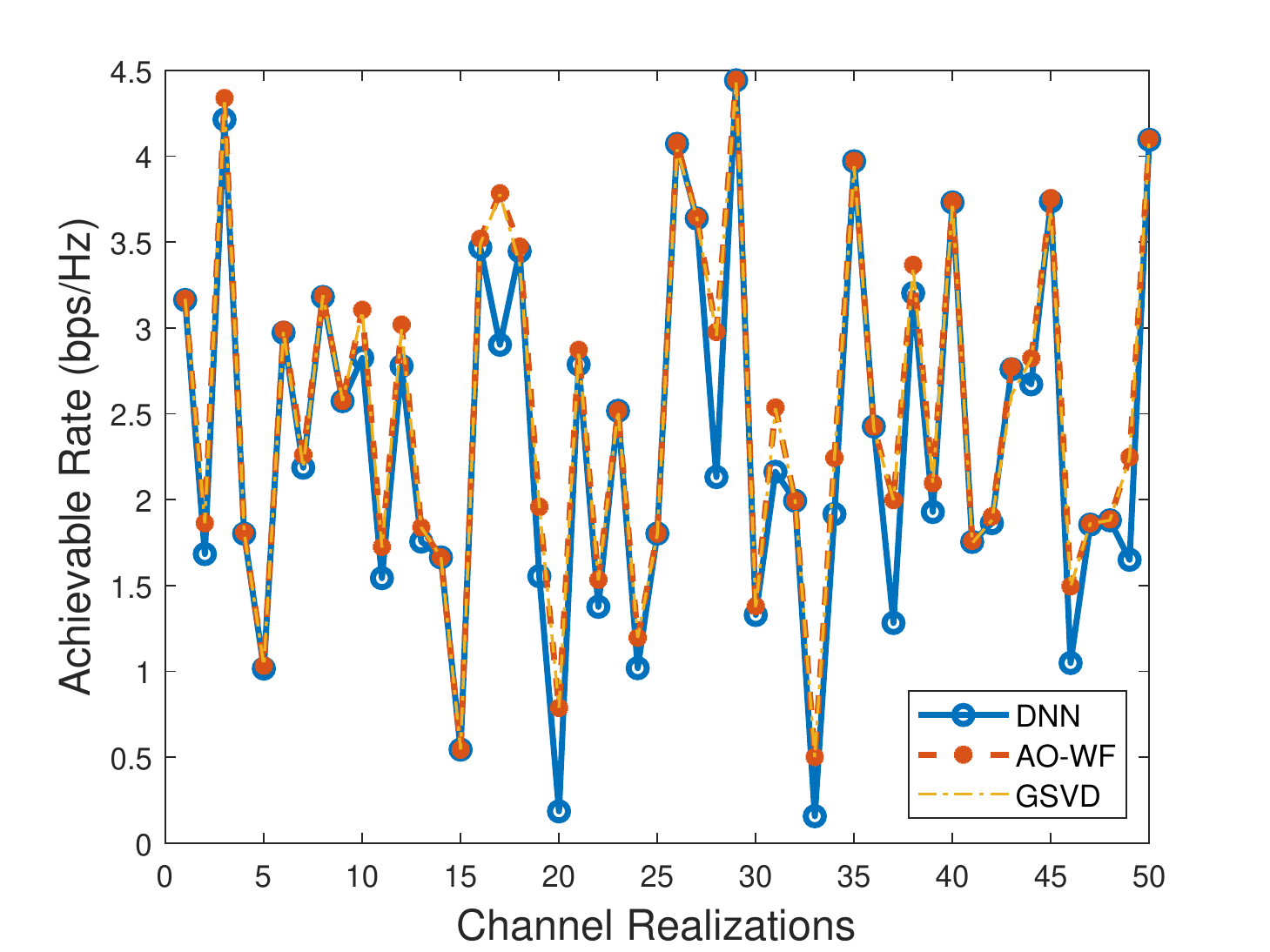}
		\label{fig:AchivaRate_jointly_343}}
	\caption{Comparison of achievable secrecy rate using combined training set and separate test sets.}
\end{figure*}

If we exchange the test and training data sets in previous 
simulation, i.e., 
if we use \textit{TrainingSet-II} 
with \textit{TestSet-I} and \textit{TrainingSet-I} with 
\textit{TestSet-II}, the DL-based precoding cannot  estimate  $ \mathbf{Q} $ 
very well as 
shown in Table~\ref{tab:testMSEvice}. 
This problem can be solved by cascading the two training sets 
 as a new one, named \textit{TrainingSet-III}. 
From 
Table~\ref{tab:testMScas},  it is seen that the performance becomes much better 
and reaches a similar level 
when training separately without additional training epochs by doubling the 
batch size, as mentioned in Subsection \ref{sec:training}.
The secrecy rates for this case are shown in 
Figs.~\ref{fig:AchivaRate_jointly_321} 
and 
\ref{fig:AchivaRate_jointly_343}. The average achievable rate is further shown 
in 
the last two rows of  Table~\ref{tab:Rcomp}. It is  seen that the 
proposed DL 
architecture is able to learn from existing optimal results with different $n_r$ 
and $n_e$ if it is trained with such samples. 
Overall, the DL-based precoding can achieve secrecy rate more efficiently 
than traditional iterative methods.
\begin{table}[htbp]
	\caption{MSE of DL-based Precoding with Opposite \\Training 
		and Test Data Sets.}
	\label{tab:testMSEvice}
	\centering
	\begin{tabular}{c|cc}
		\hline
		Training Data Set & 		\textit{TrainingSet-II} & 	
		\textit{TrainingSet-I} \\
		Test Data Set  & \textit{TestSet-I} & 	\textit{TestSet-II} \\ 
		\hline
		MSE of 	$\hat{q}_{11}$ & 2.8353  & 7.4462                                 \\
		MSE of	$\hat{q}_{22}$ & 2.8124 & 7.7543                                  \\
		MSE of	$\hat{q}_{33}$  & 3.0579 & 6.8545                                  \\
		MSE of	$\hat{q}_{12}$ & 2.7646 & 5.4432                                   \\
		MSE of	$\hat{q}_{23}$ & 2.3103 & 4.4098                                   \\
		MSE of	$\hat{q}_{13}$& 2.1800  & 5.0406                                   \\ 
		\hline
		
	\end{tabular}
\end{table}	
\begin{table}[htbp]
	\caption{MSE of DL-based Precoding with Cascaded Training Sets.}
	\label{tab:testMScas}
	\centering
	\begin{tabular}{c|cc}
		\hline
		Training Data Set & 		\multicolumn{2}{c}{\textit{TrainingSet-III}} \\
		Test Data Set  & \textit{TestSet-I} & 	\textit{TestSet-II} \\ 
		\hline
		MSE of 	$\hat{q}_{11}$ & 0.1313  & 0.1564                                 \\
		MSE of	$\hat{q}_{22}$ & 0.1234  & 0.1386                                  \\
		MSE of	$\hat{q}_{33}$ & 0.1219  & 0.1364                                    \\
		MSE of	$\hat{q}_{12}$ & 0.1526  & 0.1351                                 \\
		MSE of	$\hat{q}_{23}$ & 0.1057  & 0.1596                                   \\
		MSE of	$\hat{q}_{13}$ & 0.1384  & 0.1123                                   \\ 
		\hline
	\end{tabular}
\end{table}	
\subsection{Applying a Deeper Network}
The performance of DL-based precoding can be further improved by 
increasing the 
depth of the NN. The network architecture in Fig.~\ref{fig:network} (denoted 
as 
\textit{DeepNet}) contains $7$ FCNN layers, $9$ PReLU activation layers, 
$4$ shortcut connections, and $4$ addition nodes. If we add anther $4$ 
FCNN layers and repeat the shortcut connections, a deeper network named 
as \textit{DeeperNet} can be obtained. The \textit{DeeperNet} contains $7$ 
FCNN layers, $17$ PReLU activation layers, $8$ shortcut connections, and $8$ 
addition nodes. The average achievable secrecy rates using different data 
sets are compared in Table~\ref{tab:RcompDeepShallow}. The secrecy rate 
obtained by the  \textit{DeeperNet} outperforms  that of  
the \textit{DeepNet}. However, since the depth of the network is doubled, 
the time 
consumption is increased to $0.0405{\rm ms}$ per channel realization, i.e., 
it takes two times the  \textit{DeepNet}. Therefore, the proposed
\textit{DeepNet} compromises between secrecy performance and time cost.

\begin{table}[htbp]
	\caption{Average of Achievable Secrecy Rate.}
	\label{tab:RcompDeepShallow}
	\centering
	\begin{tabular}{cc|ccc}
		\hline
		Training Set & Test Set& \textit{DeepNet} & \textit{DeeperNet} & 
		AO-WF  \\ \hline 
		\textit{TrainingSet-I} &\textit{TestSet-I}  & 3.3980& 3.4215 & 3.4741 \\
		\textit{TrainingSet-II} &\textit{TestSet-II} & 2.3381 & 2.4137 & 2.4827 \\ 
		\hline
	\end{tabular}
\end{table}

	\section{Conclusions}\label{sec:conclu}
	In this paper, a DL-based precoding  has been proposed for the MIMO Gaussian 
	wiretap  channel. The input features of the DL-based precoding are 
	 generated by channel matrices and the output have the elements of 
	 the   covariance matrix. 
	 The network  is build up with FCNN, residual connections, 
	 and PReLU. The 
	 experiments show that the proposed precoding is much  faster than 
	 existing methods and achieves a reasonable and stable 
	 secrecy performance.  The  method is 	 energy-saving and much less  complex which makes it a promising approach to physical layer security of IoT networks.

	One practical  issue in the context of the wiretap channel is that the number of antennas at the
	eavesdropper is assumed  known. This work makes  meaningful progress toward eliminating or, at least, being less dependent on this assumption.
%	In practical scenarios, it is reasonable to assume we the 
%	 number of  antennas at  the transmitter and receiver side are known. Then 
%	 we can build data sets for training process. 
%	  The developed network is trained and tested for a fixed transmit power is fixed as 
%	 20W, which is not  flexible enough for practical usages. These are the 
%	 promising problems  worth further studies. 

\end{document}